# Thermo-mechanically coupled phase-field fracture model considering elastocaloric effect of shape memory alloy

Shen Sun[1], Wei Tang[1,2], Weiwei He[1,3], Igor Polozov[4], Min Yi[1,*]

[1] *State Key Laboratory of Mechanics and Control for Aerospace Structures & Key Laboratory for Intelligent Nano Materials and Devices of Ministry of Education &*
*Institute for Frontier Science & College of Aerospace Engineering, Nanjing University of Aeronautics and Astronautics (NUAA), Nanjing 210016, China*
[2] *IMT School for Advanced Studies Lucca, Piazza San Francesco 19, Lucca 55100, Italy*
[3] *Instituto de Ciencia de Materiales de Madrid (ICMM-CSIC), Madrid 28049, Spain*
[4] *Institute of Mechanical Engineering, Materials, and Transport, Peter the Great St. Petersburg Polytechnic University, Polytechnicheskaya 29, St. Petersburg 195251, Russia*
\* yimin@nuaa.edu.cn

**Abstract**
Modelling fracture behavior of the shape memory alloy (SMA) that interacts with martensitic transformation and the associated elastocaloric effect (eCE) still remains challenging. Herein, a thermo-mechanically coupled phase-filed fracture model considering elastocaloric effect of SMA is proposed to simulate the cracking process coupled with the non-isothermal martensitic transformation and the associated eCE. In the phase-field model, both the thermal strain induced by eCE and the eigen strain induced by the phase transition are considered. An empirical degradation function is adopted to describe the thermal conductivity decreasing with the fracture order parameter. The model is validated with the finite element method and tensile fracture properties of Mn-Cu SMA are simulated. It is found that the martensite variant nucleates at the stress concentration where the crack initiates, and commonly spreads with an angle of 45 degree. The thermal expansion strain caused by the eCE could strengthen the critical load capacity. A large kinetic parameter for phase transition and the large orientation angle could enhance the strength and temperature change of eCE while the deformation capacity is reduced. The phase-field model demonstrates its ability in the thermal-mechanically coupled toughening of SMA. It also provides a possible fracture-resistance strategy by the utilization of eCE for elastocaloric devices.

## 1 Introduction

Owing to the extraordinary shape memory response and elastocaloric effect, shape memory alloys have been employed as a kind of significant functional materials for decades, especially in aerospace [1-3], micro-electro-mechanical system [4-6], robot system [7-9], petroleum industry [10-12], biomedical field [13-15], and so on. Thermal and mechanical loads could induce the martensite phase transition or its reverse. However, the complicated stress state and deformation mode in serve may damage the shape memory alloys. The martensite phase transition response of the cracked shape memory alloy would be restricted seriously. At the same time, compared to the regular materials, the eigen strain tensor introduced by the martensite variants could affect the fracture behavior of the shape memory alloys. Thus, it is highly required to investigate the multi-physical complex fracture issue of shape memory alloys [16, 17].

The traditional fracture analysis of shape memory alloy combines the phenomenal constitution of shape memory alloy with the failure criterion in fracture mechanism. It macroscopically reveals the fracture behavior of shape memory alloys and qualitatively observes the microstructure [18-22] while the microscopic mechanism between the martensite phase transition and the crack topology still needs to be clarified further. The phase-field model is an appropriate method to handle this kind of multi-physical problems. It has great advantages in simulating the martensite variants microstructure evolution and the smear crack propagation at the same time. On the one hand, martensitic variants in martensitic transformation are taken as the order parameters in the phase-field model [23-26]. The so-called Allen-Cahn equation was applied to govern the evolution of each variant. The phase transformation types covered the hexagonal-to-orthorhombic [27], cubic-to-tetragonal [28], and tetragonal-to-monoclinic [29] transitions. Further, the thermal conduction issue could be added into the investigations of shape memory alloys. The non-isothermal phase-field model could study the elastocaloric effect and the latent heat in shape memory alloys [30-32]. For taking the most use of the elastocaloric effect, numerous efforts have been made to improve the adiabatic temperature change [33-37]. On the other hand, the phase-field method treats the sharp crack into a smeared crack region. A dimensionless order parameter is defined to describe the state of fracture. Francfort and Marigo [38] first proposed a phase-field fracture model, and then Bourdin et al. [39] numerically completed the model for quasi-static brittle fracture. For a precise description of fracture driving force, scholars made great tempts on the split of strain energy. Amor et al. [40] proposed the volumetric-deviatoric decomposition of the elastic strain tensor while Miehe et al. [41] elaborated the spectral split on the strain tensor to distinguish the elastic strain energy density. Based on the concept of structured deformations, Freddi and Royer-Carfagni [42] developed the no-tension split to describe the tension-compression asymmetry in the stiffness degradation response. Due to the flexibility of the method, the phase-field fracture model is further validated for various material properties and complex multi-physical scenes, such as material anisotropy [43, 44], heterogeneity [45, 46], elastoplasticity [47, 48], thermal-mechanically coupled fracture [49-51], hydraulic fracture [52, 53], fatigue fracture [54-56] and so on.

In order to clarify the fracture behavior of shape memory alloys, some researchers have developed phase-field models to simulate the evolution process of martensitic transformation and crack propagation. Zhu and Luo [57] studied the microcrack nucleation induced by phase transition. Sun et al. [58] analyzed the toughening effect of phase transition on crack propagation. Cissé and Zaeem [59] investigated the coupling effect between crack evolution and phase transition under a constant initial temperature. While the single physical load has been analyzed, the complex coupling between phase transition and crack evolution is not explored at the same time. Simoes and Martínez-Paneda [60] firstly proposed the phase-field model applicable to the fracture behavior of thermally induced shape memory alloys. The order parameter is defined as the percentage of martensitic phase and the different variants are not distinguished. Hasan et al. [61] developed a finite-strain phase-field fracture model which could describe both the transformation and reorientation toughening behavior of shape memory alloys. Xiong et al. [62] proposed a non-isothermal phase-field model to predict the temperature-dependent transformation and reoriention toughening behaviors of NiTi shape memory alloys.

In this work, the phase-field model combines the non-isothermal martensite phase transition with quasi-static fracture. The coupling effects between thermal conductivity degradation, martensitic transformation and crack propagation are probed. The article is organized as follows. In Section 2, the thermo-mechanically coupled phase-field fracture model is presented, with a consideration of martensite phase evolution. Section 3 illustrates the numerical examples on fracture of shape memory alloys under various load conditions. Section 4 summarizes the work.

## 2 Thermo-mechanically Coupled Phase-field Fracture Model

In this section, a thermo-mechanically coupled phase-field fracture model of shape memory alloy is proposed. Non-conserved order parameters $\eta_i$ ($i$=1,2,3) are defined to represent the martensite phases while fracture order parameter $c$ describes the crack topology.

2.1 Free energy density

The total free energy density consists of the chemical energy density, gradient energy density, elastic strain energy density and fracture energy density, e.g.,

$$\psi_{\text{total}} = \psi_{\text{che}} + \psi_{\text{gra}} + \psi_{\text{ela}} + \psi_{\text{fra}}.$$

A Landau 2-3-4 polynomial is adapted to expressed the chemical energy as

$$\psi_{\text{che}} = \frac{A}{2}(\eta_1^2 + \eta_2^2 + \eta_3^2) - \frac{B}{3}(\eta_1^3 + \eta_2^3 + \eta_3^3) + \frac{C}{4}(\eta_1^2 + \eta_2^2 + \eta_3^2)^2,$$

where $A = 32\Delta G^*$, $B = 3A - 12\Delta G_{\text{m}}$, $C = 2A - 12\Delta G_{\text{m}}$ are positive temperature-dependent coefficients. $\Delta G^*$ is the temperature-dependent energy barrier between austenite and martensite which is related to the latent heat of transformation and the energy required for overcoming interfacial and elastic distortions during the transformation. The temperature dependency of $\Delta G^*$ is expressed as

$$\Delta G^* = \begin{cases} \dfrac{0.3}{32} Q, T \leq T_0 \\ \dfrac{0.8 + 0.06(T - T_0)}{32} Q, T > T_0 \end{cases}$$

$\Delta G_{\text{m}}$ is the driving force of martensite transformation. It is a continuous function related to Clausius-Clapeyron equation and can be expressed as $\Delta G_{\text{m}} = \frac{Q(T-T_0)}{T_0}$. $Q$ is the specific latent heat and $T_0$ is the chemical equilibrium temperature. The gradient energy density at the martensite variants interface can be expressed as

$$\psi_{\text{gra}} = \frac{1}{2}\beta \sum_{i=1}^{3}(\nabla \eta_i)^2,$$

where $\beta$ is the gradient energy coefficient related to $\Delta G^*$. The elastic strain energy is considered to be the function of elastic strain and fracture order parameter. The constitution of shape memory alloy in this paper is supposed to be isotropic. The spectral split of strain tensor is applied to derived the elastic energy of cracked materials. The elastic energy density is given as

$$\psi_{\text{ela}} = g(c)\psi_{\text{ela}+} + \psi_{\text{ela}-},$$

$$\psi_{\text{ela}+} = \frac{\lambda}{2}\left[\langle \text{tr}(\varepsilon^{\text{ela}})\rangle_+\right] + \mu \text{tr}\left[\left(\varepsilon^{\text{ela}+}\right)^2\right],$$

$$\psi_{\text{ela}-} = \frac{\lambda}{2}\left[\langle \text{tr}(\varepsilon^{\text{ela}})\rangle_-\right] + \mu \text{tr}\left[\left(\varepsilon^{\text{ela}-}\right)^2\right]$$

where $\lambda$ and $\mu$ are Lame constants. The operators $\langle \cdot \rangle_+$ and $\langle \cdot \rangle_-$ are defined as $\langle \cdot \rangle_+ = (\cdot + |\cdot|)/2$ and $\langle \cdot \rangle_- = (\cdot - |\cdot|)/2$. $g(c) = (1-c)^2$ is the fracture degradation function and the typical quadratic mode is adopted. The decomposition is based on the positive and negative components of the elastic strain tensor. Let

$$\varepsilon^{\text{ela}} = \mathbf{P}\mathbf{\Lambda}\mathbf{P}^{\text{T}},$$

where $\mathbf{P}$ consists of the orthonormal eigenvectors of $\varepsilon^{\text{ela}}$ and $\Lambda = \text{diag}(\lambda_1, \lambda_2, \lambda_3)$ is a diagonal matrix of principal elastic strains. It can be defined that

$$\varepsilon^{\text{ela}+} = \mathbf{P}\mathbf{\Lambda}^+\mathbf{P}^{\text{T}},$$

$$\varepsilon^{\text{ela-}} = \mathbf{P}\mathbf{\Lambda}^{-}\mathbf{P}^{\text{T}},$$

where

$$\mathbf{\Lambda}^{+} = \text{diag}(\langle\lambda_1\rangle_{+},\langle\lambda_2\rangle_{+},\langle\lambda_3\rangle_{+}),$$
$$\mathbf{\Lambda}^{-} = \mathbf{\Lambda} - \mathbf{\Lambda}^{+}.$$

The fracture energy density is

$$\psi_{\text{fra}} = \frac{G_c}{2l_0}\left(c^2 + l_0^2\|\nabla c\|^2\right),$$

where $G_c$ is the critical fracture energy release and $l_0$ represents the surface width of the smear crack topology. $G_c$ is supposed to be the function of temperature. Assuming that the critical fracture energy release under a high temperature will be reduced [63], the temperature-dependent function is given as

$$G_c(T) = G_{c0}\left[1 - b_1\frac{T - T_{\text{ref}}}{T_{\max}} + b_2\left(\frac{T - T_{\text{ref}}}{T_{\max}}\right)^2\right].$$

Here, $b_1$ and $b_2$ are model parameters. $T_{\text{ref}}$ and $T_{\max}$ are the reference temperature and the maximum temperature, respectively. $G_{c0}$ is the value of critical fracture energy release at $T_{\text{ref}}$.

2.2 Governing Equations

The field equilibrium equations in this phase-field model conclude the mechanical stress balance law, thermal conduction equation and two Allen-Cahn equations for order parameters. The quasi-static stress balance equation and boundary conditions can be given as

$$\sigma_{ij,j} + f_i = 0 \quad \text{in } \Omega,$$

$$u_i = \hat{u}_i \quad \text{on } \partial\Omega_u, \quad \sigma_{ij}n_j = \hat{t}_i \quad \text{on } \partial\Omega_\sigma,$$

where $\sigma_{ij}$ is the stress tensor and $f_i$ is the body force. If the body force is not considered, $f_i$ is set to be zero. $\hat{u}_i$ and $\hat{t}_i$ are the displacement and surface traction on the boundary part $\partial\Omega_u$ and $\partial\Omega_\sigma$, respectively. For brittle shape memory alloys, the linear elastic constitution is adopted, e.g.,

$$\sigma_{ij} = C_{ijkl}\varepsilon_{kl}^{\text{ela}},$$

where $C_{ijkl}$ is the stiffness matrix. The total strain tensor of the material includes the thermal strain, elastic strain and the eigen strain induced by the martensitic transformation, e.g.,

$$\varepsilon_{ij}^{\text{total}} = \varepsilon_{ij}^{\text{th}} + \varepsilon_{ij}^{\text{ela}} + \varepsilon_{ij}^{\text{eigen}}.$$

The thermal strain is described as

$$\varepsilon_{ij}^{\text{th}} = \alpha_T(T - T_0)\delta_{ij},$$

where $\alpha_T$ is the thermal expansion coefficient. $T_0$ is the initial temperature without thermal strain which is set to be the same as the chemical equilibrium temperature. The eigen strain tensor is defined as

$$\varepsilon_{ij}^{\text{eigen}} = \varepsilon_{ij}^{00}(1)\eta_1 + \varepsilon_{ij}^{00}(2)\eta_2 + \varepsilon_{ij}^{00}(3)\eta_3.$$

The tetragonal transformation strain is described with

$$\varepsilon_{ij}^{00}(1) = \begin{pmatrix} \varepsilon_3 & 0 & 0 \\ 0 & \varepsilon_1 & 0 \\ 0 & 0 & \varepsilon_1 \end{pmatrix}, \quad \varepsilon_{ij}^{00}(2) = \begin{pmatrix} \varepsilon_1 & 0 & 0 \\ 0 & \varepsilon_3 & 0 \\ 0 & 0 & \varepsilon_1 \end{pmatrix}, \quad \varepsilon_{ij}^{00}(3) = \begin{pmatrix} \varepsilon_1 & 0 & 0 \\ 0 & \varepsilon_1 & 0 \\ 0 & 0 & \varepsilon_3 \end{pmatrix}.$$

Due to the switching of crystal orientation with Cartesian coordinates, the angle $\theta$ is set to be between the a-axis of the tetragonal cell and the x-axis of the global coordinate system. The 2D rotation matrix is written as

$$R_{ij}(\theta) = \begin{pmatrix} cos(\theta) & sin(\theta) \\ -sin(\theta) & cos(\theta) \end{pmatrix}.$$

Considering the non-isothermal process, the evolution of temperature can be derived as

$$\rho c_p \frac{\partial T}{\partial t} = \nabla(\kappa \nabla T) + Q \sum_{i=1}^{3} \eta_i^2,$$

where $\rho$ is the material density and $c_p$ is the specific heat. $\kappa$ is the thermal conductivity which is assumed to degrade dependent on the crack order parameter, i.e., $\kappa = \kappa_0 m(c)$. Raised by Svolos et al. [50], the thermal conductivity degradation function is given as

$$m(c) = \frac{1 - \left[\sqrt{c(2-c)}\right]^3}{1 + \frac{1}{2}\left[\sqrt{c(2-c)}\right]^3}.$$

The evolution of martensite variants $\eta_i$ is controlled by the time-dependent Ginzburg-Landau equation, i.e.,

$$L\dot{\eta}_i = -\frac{\delta \psi}{\delta \eta_i},$$

where $L$ is the kinetic parameter characterizing the interfacial migration. Another Allen-Cahn equation for the fracture order parameter $c$ is given as

$$\gamma \dot{c} = -\frac{\delta \psi}{\delta c},$$

where $\gamma$ is also a viscosity parameter. $L$ and $\gamma$ are key coefficients which control the evolution speeds of order parameters. The relationship between their numerical values critically affects the simulation results. To simplify the analysis, $\gamma$ is set to be zero and the quasi-static fracture is assumed. The influence of $L$ on the fracture of shape memory alloys will be discussed in the following. Combining the total free energy density function, the evolution of $c$ can be written as

$$\frac{G_c}{l_0} c + 2(1-c)\psi_{\text{ela}+} + G_c l_0 \nabla^2 c = 0.$$

Considering that there is no crack healing behaviour in the simulation, a history variable $\mathcal{H}(t)$ is introduced to prevent the fracture order parameter $c$ from decreasing. It is given as

$$\mathcal{H}(t) = \max_{s \in [0,t]} \psi_{\text{ela}+}(s).$$

The controlling equation of $c$ is finally extended to be

$$\frac{G_c}{l_0} c + 2(1-c)\mathcal{H}(t) + G_c l_0 \nabla^2 c = 0.$$

**3 Results and discussion**

The phase-field model is solved by finite element method (FEM). The governing equations are discretized spatially and temporally. Method of weighted residuals is used to solve the equations. The FEM programming and calculation are performed in the open-source Multiphysics Object Oriented Simulation Environment (MOOSE). Mn-Cu alloy is adopted in the simulations and the material parameters are summarized in Table 1. Due to the negligible differences between austenite and martensite in Mn-Cu alloy, the parameters are assumed to be the same for different phases. To simplify the numerical complexity, a two-dimensional domain is utilized in the calculation. Two martensite variants in the plane are considered and the initial phase of the plate is austenite. Taking the phase transition interface and crack interface into consideration, the minimum mesh size is set to be 10 nm. The fracture processed of single crystal and bicrystal are analyzed respectively in the following.

3.1 Single-edge notched tension of single crystal

A squared plate with a horizontal notch which is placed at middle height from the left outer surface to the center of the specimen is simulated in this subsection. The plate size is set as 1000 nm× 1000 nm and the notch is 500 nm. The total number of the meshes is 53207 which can make the phase-field interface cover at least two elements. A tensile load is applied on the top of the specimen while the bottom of the specimen is fixed, as shown in Fig. 1(a). All the boundaries are adiabatic. The crystal orientation is in the horizontal direction.

Table 1 Material parameters of Mn-Cu shape memory alloy

| Parameter | Value |
| --- | --- |
| Latent heat $Q$ (J/m3) | $4.84\times10^7$ |
| Chemical equilibrium temperature $T_0$ (K) | 245 |
| Gradient energy coefficient $\beta$ (J/m) | $2.5\times10^{-9}$ |
| Kinetic parameter $L$ (kg/m/s) | 0.02 |
| $\varepsilon_1$ | -0.01 |
| $\varepsilon_3$ | 0.01 |
| Density $\rho$ (kg/m3) | $7.5\times10^3$ |
| Specific heat $c_p$ (J/kg/K) | 352 |
| Thermal conductivity $\kappa_0$ (W/m/K) | 40 |
| Thermal expansion coefficient $\alpha_T$ (1/K) | $10^{-5}$ |
| Lame constant $\lambda$ (GPa) | 14.588 |
| Lame constant $\mu$ (GPa) | 31 |
| Critical fracture energy release $G_{c0}$ (N/m) | 0.8 |
| Length scale $l_0$ (nm) | 10 |
| $b_1$ | 1.8 |
| $b_2$ | 1.1 |
| Reference temperature $T_{\text{ref}}$ (K) | 293 |
| Maximum temperature $T_{\max}$ (K) | $10^3$ |

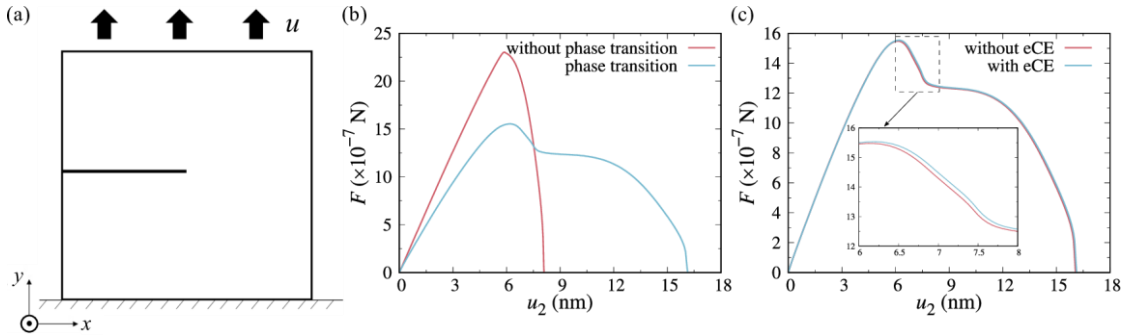

Fig.1. (a) Geometry and boundary conditions of the single-edge notch tension. (b) Load-displacement curves of the single-edge notch tension with/without phase transition. (c) Load-displacement curves of the single-edge notch tension with/without eCE

The displacement-load curves are shown in Fig. 1(b). A simulation without phase transition is applied for comparison. Except for not taking the eigen strain and martensite phase transition into account, other material parameters are the same as those of the Mn-Cu shape memory alloy. The Mn-Cu alloy has a lower critical load while it could bear larger deformation. The patterns of martensite phase transition and crack propagation process are plotted in Fig. 2(a-c) and the temperature distributions are shown in Fig. 2(d-f). Around the notch tip, the stress concentration induces the austenite to transform into martensite. The tensile load in the vertical direction makes $\eta_1$ evolve to decrease the elastic strain $\varepsilon_{22}$. Martensite variant I nucleates along the 45° direction relative to the horizontal direction. The crack initiates at the tip of the notch and propagates horizontally. During the phase transition and crack propagation, the temperature around the phase transition region rises and the maximum temperature change is about 9 K.

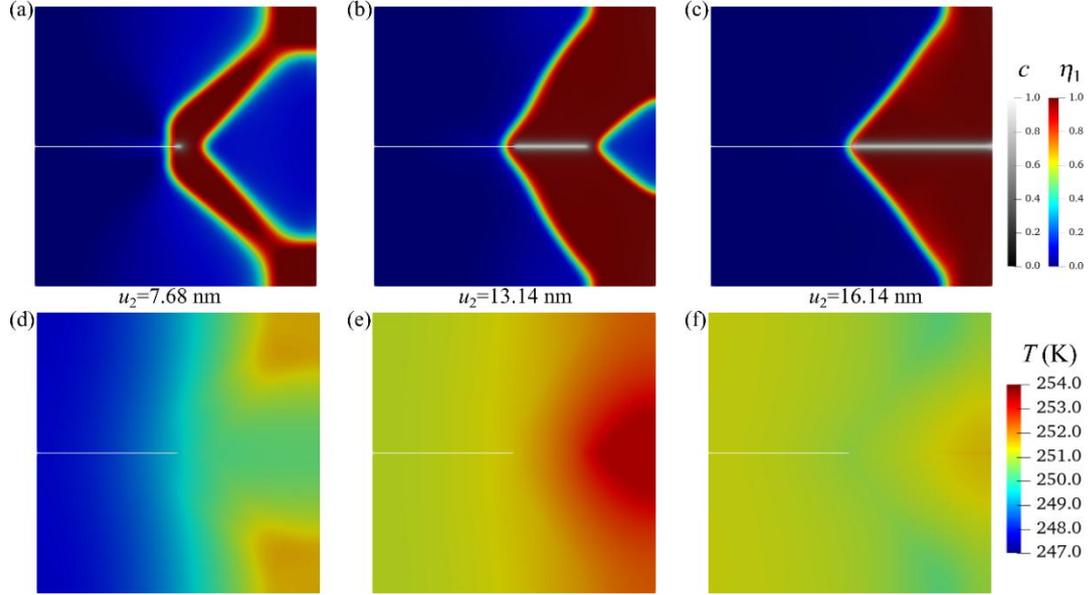

Fig. 2 Martensite phase patterns and crack propagation under the displacement of (a) $u_2$=7.68 nm, (b) $u_2$=13.14 nm and (c) $u_2$=16.14 nm. Temperature distribution under the displacement of (d) $u_2$=7.68 nm, (e) $u_2$=13.14 nm and (f) $u_2$=16.14 nm.

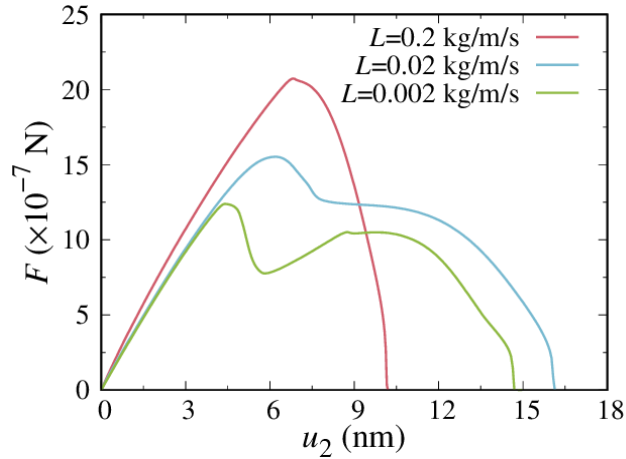

Fig. 3 Load-displacement curves of the single-edge notch tension under different kinetic parameters

Fig. 1(c) shows the load-displacement curves of single-edge notch tension with and without eCE. During the loading stage, the two curves have no difference. In the crack propagation stage, the load without eCE is slightly smaller than that with eCE. Due to the thermal expansion strain caused by the increased temperature, the elastic strain could be partially compensated. The critical load is increased marginally and the crack propagation speed is delayed.

3.1.1 Effects of the kinetic parameter $L$

In order to analyze the influence of $L$ on the phase transition rate, the kinetic parameter is additionally set to be 0.002 and 0.2, respectively. Other parameters and boundaries remain unchanged. The kinetic parameter controls the evolution speed of $\eta_i$. A large kinetic parameter $L$ leads to a slow evolution of $\eta_i$. The displacement-load curves under different kinetic parameters are depicted in Fig.3. The martensite variant structures and the crack patterns are shown in Fig.4. Due to the retardation of large $L$, the degree of phase transition is quite low. The large $L$ causes a high critical load which is similar to the single-edge notch tension result without phase transition. A small $L$ makes the phase transition in a large aera around the crack tip of the specimen. The eigen strain caused by the $\eta_1$ drives the crack propagates

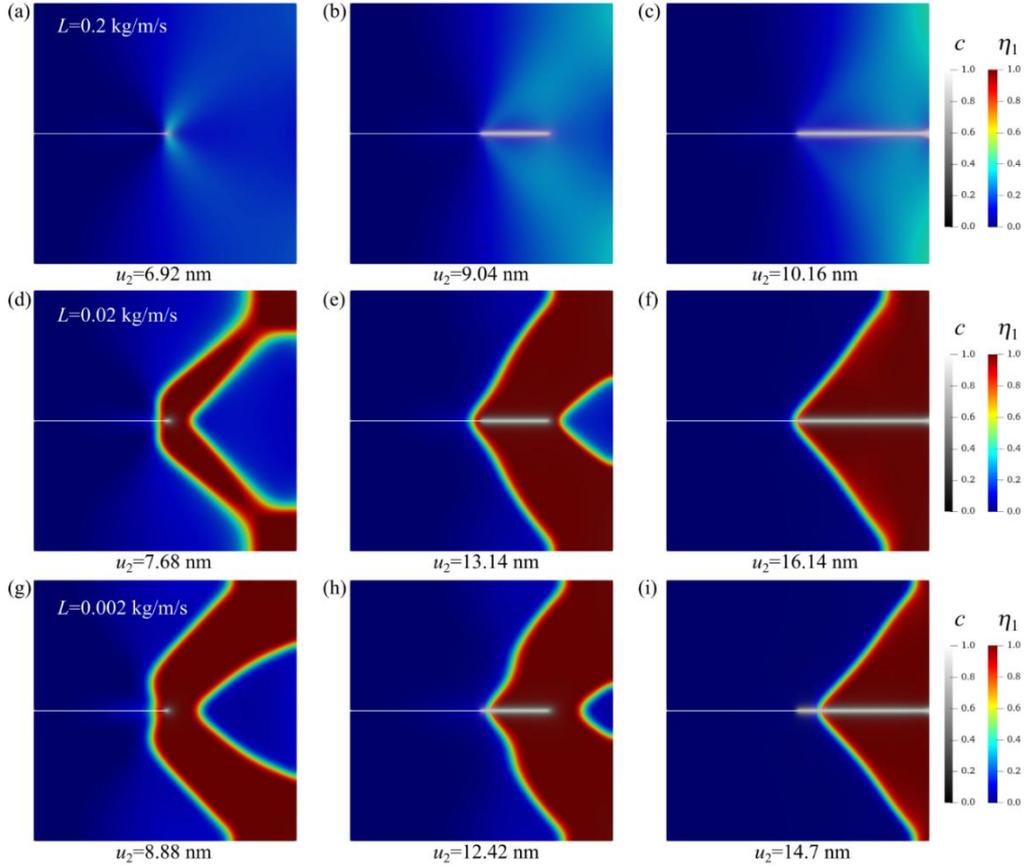

Fig. 4 Martensite phase patterns and crack propagation under the kinetic parameter of (a-c) $L = 0.2\ \mathrm{kg/m/s}$, (d-f) $L = 0.02\ \mathrm{kg/m/s}$ and (g-i) $L = 0.002\ \mathrm{kg/m/s}$.

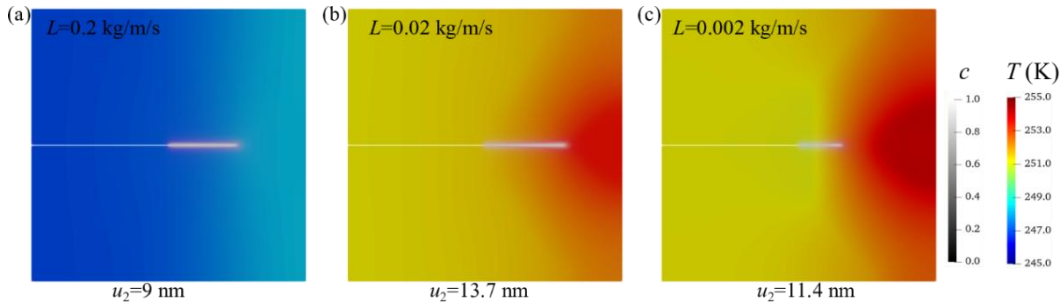

Fig. 5 (a) Temperature distribution and crack morphology at a displacement of $u_2 = 9\ \mathrm{nm}$ with $L = 0.2\ \mathrm{kg/m/s}$, (b) Temperature distribution and crack morphology at a displacement of $u_2 = 13.7\ \mathrm{nm}$ with $L = 0.02\ \mathrm{kg/m/s}$, (c) Temperature distribution and crack morphology at a displacement of $u_2 = 11.4\ \mathrm{nm}$ with $L = 0.002\ \mathrm{kg/m/s}$.

under a small load. The quasi-static fracture is considered in our work, so the evolution of fracture order parameter $c$ is time-independent and instantaneous. When the crack cross over the phase transition region, the load curve rises again. The crack continues to move on until the phase transition on the new crack tip evolves. Then there is a second drop in the displacement-load curve. Due to the difference of the kinetic coefficients, the percentages of martensitic transformation in the specimens are various. Under the coupling of mechanical loading and phase transition, the temperature change inside the single edge

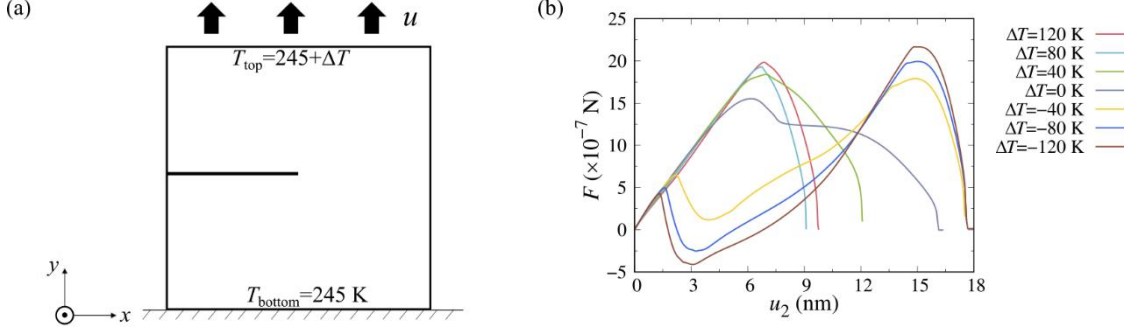

Fig. 6 (a) Boundary conditions of the single-edge notch tension with thermal loads. (b) Load-displacement curves under different thermal loads.

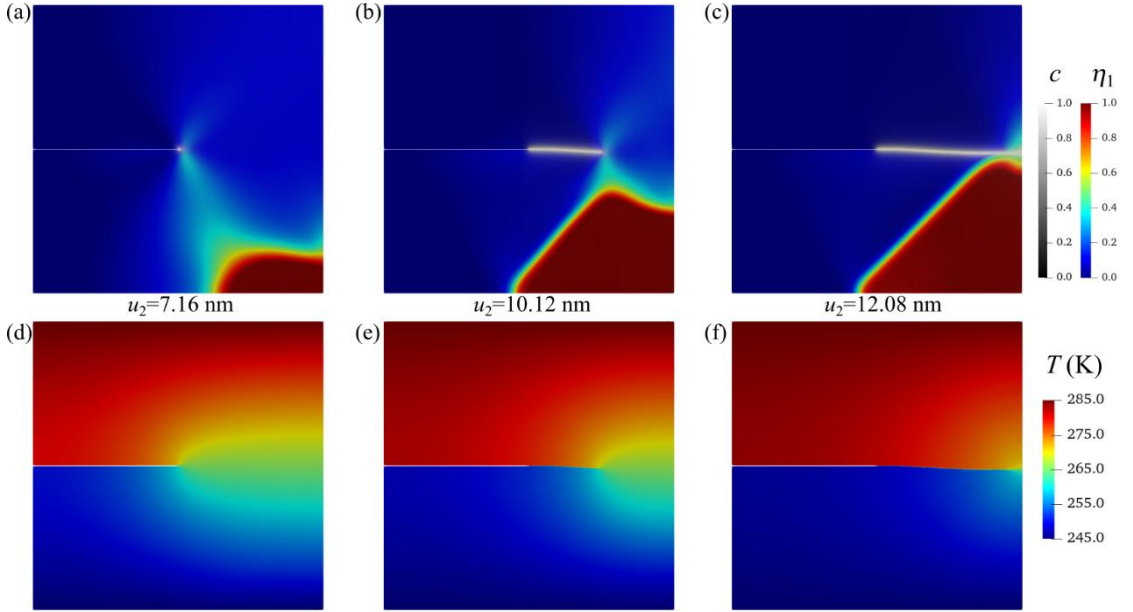

Fig. 7 Martensite phase patterns, crack propagation and temperature distributions with $\Delta T = 40$ K.

notched plate differs. Fig.5 illustrates the temperature distributions and crack propagation patterns at the moment of maximum temperature change for different kinetic coefficients. As the kinetic coefficient $L$ decreases, the extensive martensitic transformation leads to a higher temperature rise in the plate. The maximum temperature occurs along the path of horizontal crack propagation. For the three kinetic coefficients $L = 0.2, 0.02$ and $0.002$ kg/m/s, the corresponding maximum temperature increments are 3.2, 8.6 and 9.1 K, respectively. A larger temperature increase will induce a bigger thermal expansion strain during the tensile loading process, which slightly increases the critical displacement of the specimen.

3.1.2 Effects of the temperature load

Besides the tension load, the temperature load is applied on the top boundary in this subsection, as shown in Fig. 6(a). The temperature changes linearly in the first 100 ns and then remain constant in the following timesteps. The temperature of the bottom boundary is set to be constant $T = 245$ K. The displacement-load curves under different temperature loads are shown in Fig. 6(b). The phase transition distributions and crack patterns under the $\Delta T = 40\ K$ are depicted in Fig. 7. Martensite variant I initially nucleates in the bottom boundary and then diffuses towards the notch tip because of the low temperature. The high temperature will decrease the critical fracture energy and causes a lower critical load. The vertically gradient fracture toughness $G_c$ makes the crack path slightly deflect upward.

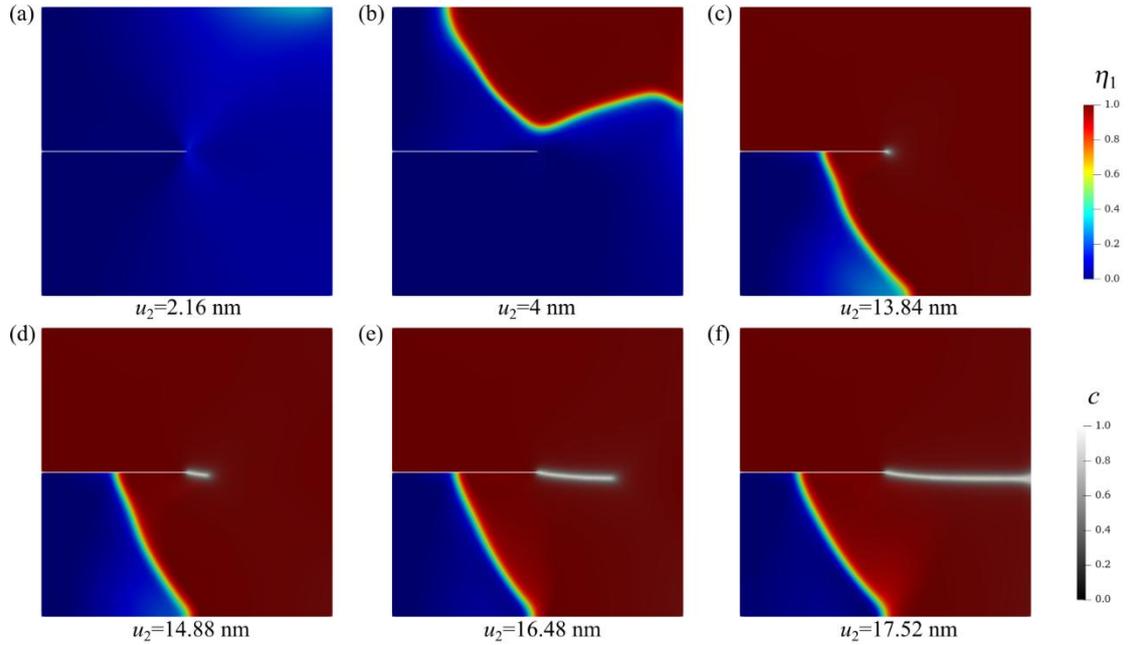

Fig. 8 Martensite phase patterns and crack propagation with $\Delta T = -40$ K under the displacement of (a) $u_2$=2.16 nm, (b) $u_2$=4 nm, (c) $u_2$=13.84 nm, (d) $u_2$=14.88 nm, (e) $u_2$=16.48 nm and (f) $u_2$=17.52 nm.

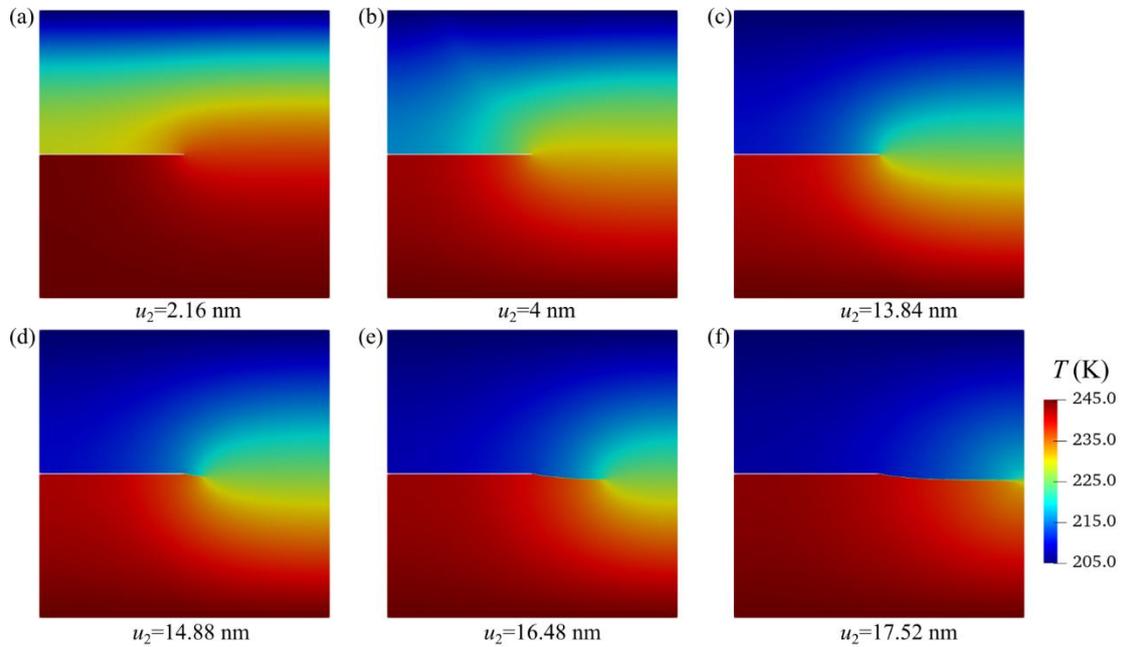

Fig. 9 Temperature Distributions with $\Delta T = -40$ K under the displacement of (a) $u_2$=2.16 nm, (b) $u_2$=4 nm, (c) $u_2$=13.84 nm, (d) $u_2$=14.88 nm, (e) $u_2$=16.48 nm and (f) $u_2$=17.52 nm.

On the other hand, the cases under decreased temperature loads are different. As shown in Fig.8, the compressive strain caused by the low temperature greatly raise the deformation durability of the Mn-Cu shape memory alloy. There are two obvious drops in the displacement-load curve. At first, the load curve rises along with the tensile displacement. Then the decreasing temperature induces compressive thermal strain and makes the load decline. Upon the displacement load is large enough to cover the compressive

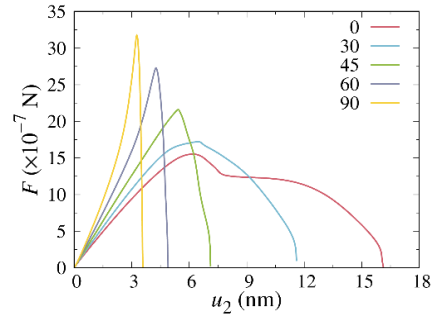

Fig. 10 Load-displacement curves under different orientation angles.

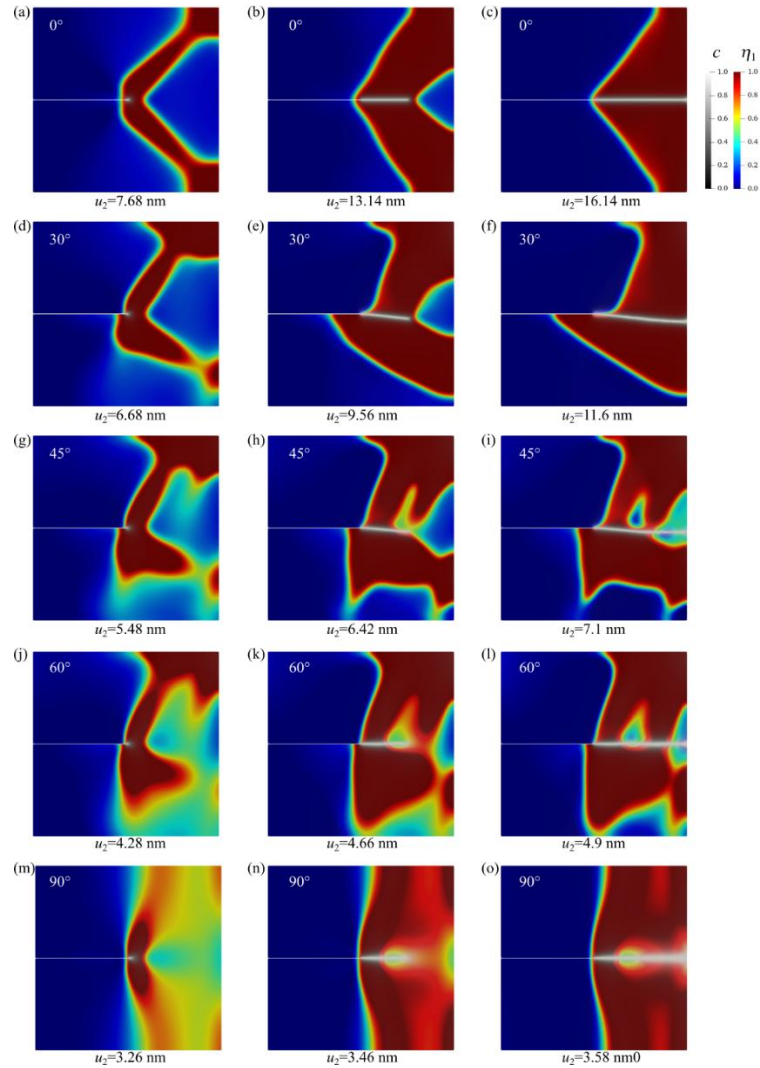

Fig. 11 Martensite phase patterns and crack propagation under the orientation angle of (a-c) 0°, (d-f) 30°, (g-i) 45°, (j-l) 60°, and (m-o) 90°.

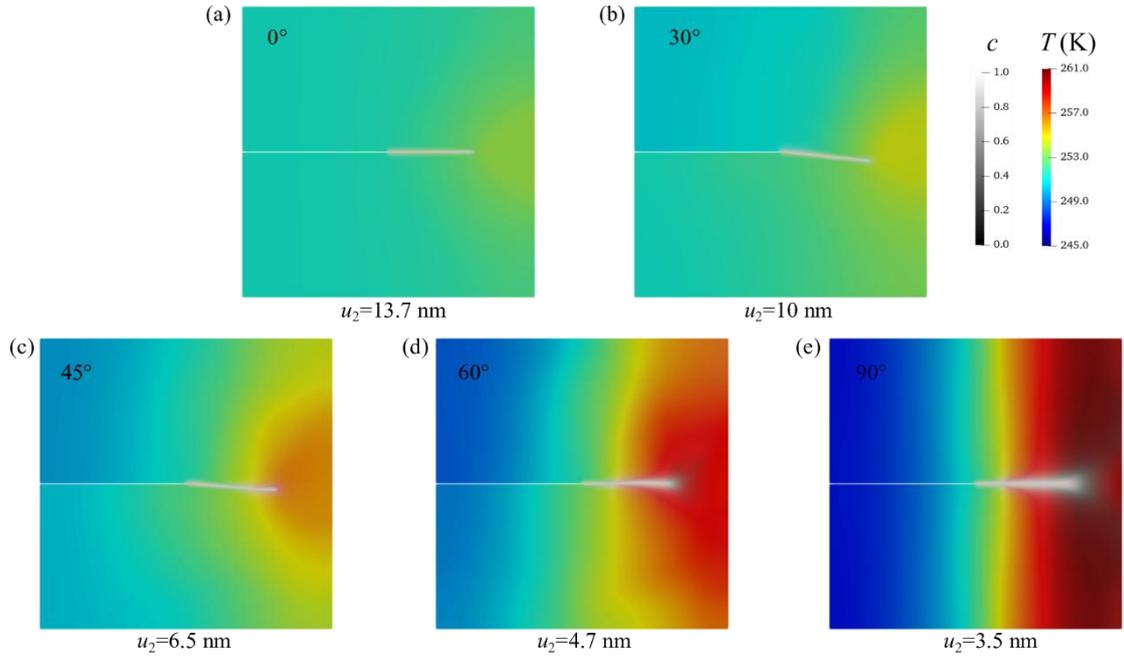

Fig. 12 (a) Temperature distribution and crack pattern with the orientation angle of 0° at $u_2 = 13.7$ nm, (b) Temperature distribution and crack pattern with the orientation angle of 30° at $u_2 = 10$ nm, (c) Temperature distribution and crack pattern with the orientation angle of 45° at $u_2 = 6.5$ nm, (d) Temperature distribution and crack pattern with the orientation angle of 60° at $u_2 = 4.7$ nm, (e) Temperature distribution and crack pattern with the orientation angle of 90° at $u_2 = 3.5$ nm.

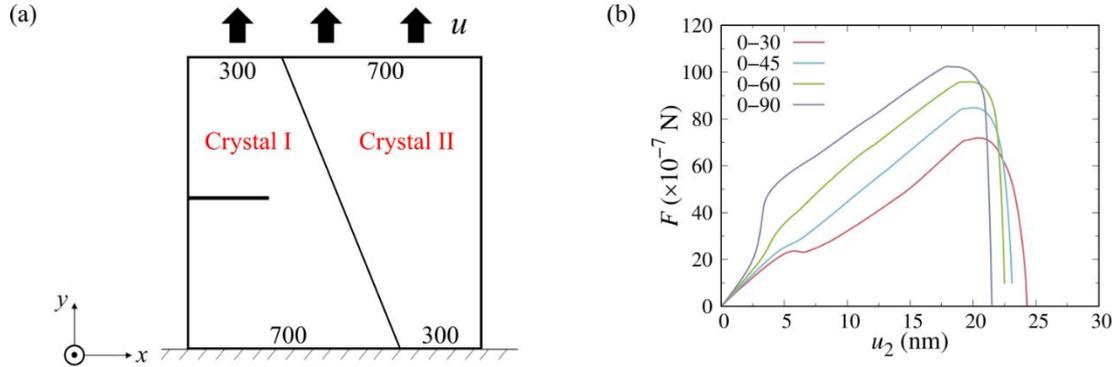

Fig. 13 Load-displacement curves of bicrystal specimen.

thermal strain, the load curve continues to increase again. When the critical load is reached, the curve finally drops. Fig.9 plots the martensite variant distributions and crack patterns under a temperature of $\Delta T = -40\,K$. The crack initiates at the notch tip and then propagates until the whole specimen failures. Because the temperature on the top boundary is decreased and is lower than the phase transition temperature, the martensite variant I nucleates near the upper right corner of the specimen. The majority aera inside the plate transforms into martensite phase $\eta_1$ before the crack nucleates.

3.1.3 Effects of the crystal orientation angle
The real crystal orientation of the shape memory alloy is various and has great impact on the fracture behaviour. 30°, 45°, 60°, 90° relative to the horizontal direction are set in this subsection. The load-displacement curves are depicted in Fig.10. A large crystal orientation angle converts the compressive eigen strain into the tensive strain. The critical load is extremely increased while the critical displacement is quite tiny. Under the angles of 30°, 45° and 60°, the martensite phase evolves along the oblique direction. 90° orientation angle drives the martensitic transformation to diffuse vertically. From the crack pattern in Fig.11, the cracks under the angle of 30° and 45° propagate slightly downward. In the simulation with the angles of 60° and 90°, the cracks are horizontal because the phase transition region is nearly symmetrical. The temperature distributions are depicted in Fig.12. A large orientation angle will bring up a high temperature change.

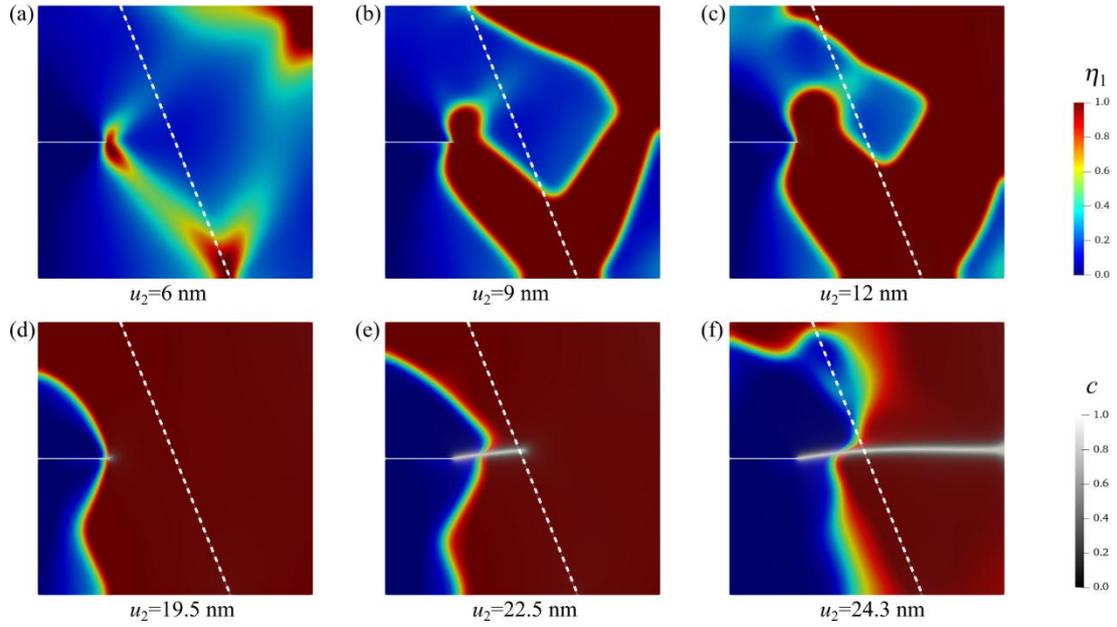

Fig. 14 Martensite phase patterns and crack propagation of the bicrystal specimen with the 0°-30° orientation angles.

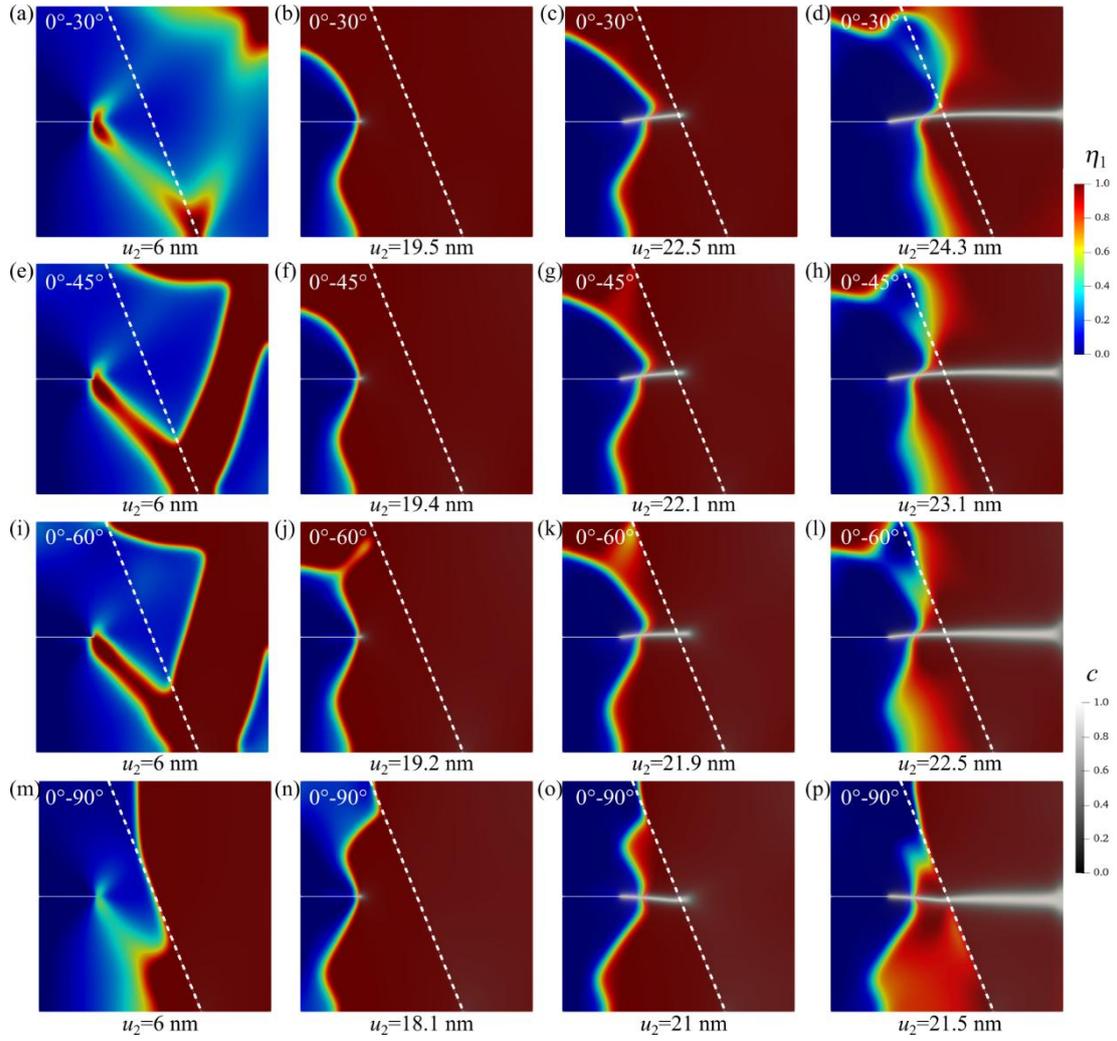

Fig. 15 Martensite phase patterns and crack propagation of the bicrystal specimen with the orientation angle of (a-d) 0°-30°, (e-h) 0°-45°, (i-l) 0°-60°, and (m-p) 0°-0°.

3.2 Single-edge notched tension of bicrystal

The specimen is changed into a bicrystal structure and the notch is shortened to 250 nm. The geometry of the plate is shown in Fig. 13(a). Two crystals are separated by a diagonal line. The left crystal is 0° while the right one is set to be 30°, 45°, 60°, and 90° in different cases, separately. To ensure the martensitic transformation occurs before the crack propagates, the critical fracture energy release is increased to 4 N/m. Load-displacement curves are depicted in Fig. 13(b). Similar to the results of single crystal, a large orientation angle follows with a higher critical load and a decreased deformation. Fig. 14 plots the martensite variant distributions and crack patterns of a 0°-30° bicrystal. The martensitic variant I initially nucleates at the bottom of the crystal boundary. Then it diffuses towards the notch tip and the right-top corner of the specimen. When the crack initiates, the majority of the plate transforms from austenite to martensite. During the propagation of the crack, the aera of the martensite phase is gradually decreasing. Fig. 15 shows the fracture processes with different orientations. The evolution speed of martensitic transformation in the large orientation angle is quite faster and the crack is more smeared.

4 Conclusion

In summary, a thermo-mechanically coupled phase-field fracture model for SMA is utilized to investigate the fracture behavior affected by martensitic phase transition and eCE. The thermal conductivity is degraded as a function of the fracture order parameter. For the consideration of crystal orientation angle, the rotation matrix is introduced into the eigen strain tensor induced by martensite variants. The phase-field fracture model is then employed to illustrate the fracture behavior of the Mn-Cu alloy. The eigen strain caused by martensite transformation greatly decreases the deformation and increased critical load of SMA. The temperature change caused by eCE induces the thermal expansion strain which retards the softening rate during the crack propagation process. A small kinetic coefficient $L$ and a large orientation angle could enhance the eCE and the temperature change thus rises. The maximum temperature change would be nearly 10 K. The non-uniform temperature distribution generated by the eCE also affects the path of crack propagation which could deflect. In a bicrystal specimen, a bigger difference in orientation angle leads to a higher critical load. The phase-field model of SMA has been demonstrated capable of simulating the fracture behavior considering the martensite transformation and eCE. The simulation results unveil the prospective strategy for improving the fracture properties of SMA by utilizing stronger eCE.

**Acknowledgments**
The authors acknowledge the support from National Natural Science Foundation of China (U2441272, 12272173, 11902150), Outstanding Youth Fund of Jiangsu Province (BK20240077), Key Project (Provincial-Municipal Joint) of Jiangsu Province (BK20243044), Fundamental Research Funds for the Central Universities (NS2023054, NE2024001), China Postdoctoral Science Foundation (2023M741690), State Key Laboratory of Mechanics and Control for Aerospace Structures (MCAS-I-0125K01), and a project Funded by the Priority Academic Program Development of Jiangsu Higher Education Institutions. This work is partially supported by High Performance Computing Platform of Nanjing University of Aeronautics and Astronautics.